\documentclass[11pt]{article}
\usepackage{amsmath,amssymb,color,graphics,epsfig,cite}

\usepackage{amsfonts}

\newcommand{\hoch}[1]{$\, ^{#1}$}

\makeatletter
\@addtoreset{equation}{section}
\makeatother

\newcommand{\ba}{\begin{eqnarray}}
\newcommand{\ea}{\end{eqnarray}}

\def\ben{\begin{equation}}
\def\een{\end{equation}}
\def\bea{\begin{eqnarray}}
\def\eea{\end{eqnarray}}
\def\be{\begin{equation}}
\def\ee{\end{equation}}

\def\nn{\nonumber}

\def\ben{\begin{equation}}
\def\een{\end{equation}}
\def\bea{\begin{eqnarray}}
\def\eea{\end{eqnarray}}
\def \nn {\nonumber}

\def\ft#1#2{{\textstyle{\frac{\scriptstyle #1}{\scriptstyle #2} } }}
\def\fft#1#2{{\frac{#1}{#2}}}

\begin{document}


\begin{center}

{\Large {\bf Bounds on Photon Spheres and Shadows of Charged Black Holes in Einstein-Gauss-Bonnet-Maxwell Gravity}}

\vspace{20pt}
{Liang Ma and  H. L\"u\hoch{*}}

\vspace{10pt}

{\it Center for Joint Quantum Studies and Department of Physics,\\
School of Science, Tianjin University, Tianjin 300350, China}

\vspace{40pt}

\underline{ABSTRACT}
\end{center}

We consider spherically symmetric and static charged black holes in Einstein-Gauss-Bonnet-Maxwell gravities in general $D\ge 5$ dimensions and study their photon spheres and black hole shadows.  We show that they all satisfy the sequence of inequalities recently proposed relating a black hole's horizon, photon sphere, shadow and its mass.

\vfill {\footnotesize liangma@tju.edu.cn, \quad mrhonglu@gmail.com}

{\footnotesize \hoch{*}Corresponding author}

\thispagestyle{empty}

\pagebreak



\newpage

\section{Introduction}

Spherically symmetric and static black holes play an important r\^ole in Einstein's General Relativity owing to their simplicity, not only for their own construction, but also for the analysis of their surrounding  geodesic motions.  The analysis \cite{Synge:1966okc,Luminet:1979nyg} of null geodesics of such a black hole that are asymptotic to the Minkowski spacetime indicates that photons can have a close orbit, forming an photon sphere. For most known exact black hole solutions, there is only one such a photon sphere and it is unstable. There are two classes of photons whose orbits do not cross the photon sphere: those inside will spiral into the horizon and those outside will escape to infinity, surrounding a shadow disk, whose radius, also called optical radius, is the impact parameter of the photons.

For spherically symmetric and static black holes, multiple close orbits can exist even under the stringent dominant energy condition and an explicit example was constructed in Einstein-Maxwell gravity extended with a quasi-topological electromagnetic structure \cite{Liu:2019rib}. In this black hole, there exists a stable photon sphere sandwiched between two unstable ones.  Thus the trapped photons inside the outer photon sphere can form a photon shield outside the horizon, without falling into the horizon or escaping to infinity.

Recently a sequence of inequalities was proposed relating the radii of the black hole event horizon $R_+$, the (outer and unstable) photon sphere $R_{\rm ph}$, the black hole shadow $R_{\rm sh}$ \cite{Lu:2019zxb}
\be
\fft32 R_+\,\le\, R_{\rm ph}\,\le\, \fft{R_{\rm sh}}{\sqrt3}\,\le\, 3M.\label{d4conjecture}
\ee
This set includes the well-known Riemann-Penrose inequality $R_+\le 2M$ \cite{Penrose:1973um}, and the inequalities proposed by Hod ($R_{\rm ph}\le 3M$) \cite{Hod:2017xkz} and Cveti\v c, Gibbons and Pope, ($R_{\rm ph}\le R_{\rm sh}/\sqrt3$) \cite{Cvetic:2016bxi}. The Riemann-Penrose inequality is considered proven under the dominant energy condition, see e.g.~\cite{Mars:2009cj}.  The other two inequalities can also be proven under the dominant energy condition, together with negative trace of the energy-momentum tensor.  However, large number of black holes satisfying at least the null energy condition were examined in \cite{Lu:2019zxb} and no counterexample was found for (\ref{d4conjecture}). A different lower bound for photon sphere was also conjectured in \cite{Hod:2012nk}, and it was shown \cite{Cvetic:2016bxi} to be violated by the Kaluza-Klein dyonic black hole \cite{Rasheed:1995zv,Lu:2013ura}.

The four-dimensional inequalities (\ref{d4conjecture}) was generalized to higher dimensions and they become
\cite{Lu:2019zxb}
\be
\big(\ft12(D-1)\big)^{\fft{1}{D-3}}R_+\,\le R_{\rm ph}\,\le
\sqrt{\ft{D-3}{D-1}}\, R_{\rm sh}\,\le\,
\Big(\ft{8\pi M(D-1)}{(D-2)\Omega_{D-2}}\Big)^{\fft{1}{D-3}}\,.\label{dconjecture}
\ee
In particular, it was stated that the Reissner-Nordstr\"om black hole in general dimensions was verified to satisfy these inequalities \cite{Lu:2019zxb}.  A sufficient energy condition for the $R_{\rm ph}$-$M$ inequality was established in \cite{Gallo:2015bda}. The purpose of this paper is not to verify the conjecture with more examples in Einstein gravity in general dimensions. Instead, we shall consider charged black holes in Einstein-Gauss-Bonnet-Maxwell (EGBM) gravities in general $D\ge 5$.  This is worth checking for two reasons.  On one hand, the theory is beyond Einstein gravity since it involves quadratic curvature invariants.  On the other hand, if we consider the Gauss-Bonnet term as matter, then the black holes satisfy the weak energy condition, which makes the verification necessary and nontrivial.

The paper is organized as follows. In section 2, we start with a review of the charged asymptotically flat black holes in EGBM gravities in general $D\ge 5$ dimensions.  We then show that the inequalities (\ref{dconjecture}) are satisfied by the simpler RN black holes and also the $D=5$ neutral black hole.  We then prove that the inequalities hold for all the static general black holes. We conclude the paper in section 3.

\section{Einstein-Gauss-Bonnet-Maxwell Gravity}

We start with the Lagrangian of EGBM gravity in general $D$ dimensions
\be
{\cal L}=\sqrt{-g} \big(R -\ft14 F^2 + \alpha_{\rm GB} E^{(4)}\big)\,,\qquad
E^{(4)}=R^2 -4 R^{\mu\nu}R_{\mu\nu} + R^{\mu\nu\rho\sigma} R_{\mu\nu\rho\sigma}.
\ee
The quadratic Euler integrand is inspired by ${\cal N}=1$ superstring, arising as an $\alpha'$ correction of the string world sheet \cite{Bergshoeff:1989de}.   The theory admits Minkowski spacetime as its vacuum and ghost-free condition requires that the coupling constant $\alpha_{\rm GB}\ge 0$ \cite{Boulware:1985wk}. The AdS vacuum of this theory, on the other hand, has ghostlike graviton modes.

\subsection{Charged black holes}

The EGBM gravity admits spherically-symmetric and static charged black holes \cite{Wiltshire:1985us,Cvetic:2001bk}, given by
\bea
ds^2_D  &=& - f dt^2 + \fft{dr^2}{f} + r^2 d\Omega_{D-2}^2\,,\qquad A =\Phi(r) dt\,,\qquad \Phi=\sqrt{\fft{2(D-2)}{D-3}}\fft{q}{r^{D-3}},\nn\\
f &=& 1 + \fft{r^2}{2\alpha}\left( 1 - \sqrt{1 + \fft{8\alpha\mu}{r^{D-1}} -
\fft{4\alpha q^2}{r^{2(D-2)}}}\right),\qquad \alpha=(D-3)(D-4)\alpha_{\rm GB}\,.\label{sol}
\eea
The mass and electric charge are given by
\be
M=\fft{(D-2)\Omega_{D-2}}{8\pi}\mu\,,\qquad Q_e = \fft{\sqrt{(D-3)(D-2)}\Omega_{D-2}}{8\sqrt2 \pi} q\,.
\ee
Here $\Omega_{D-2}$ denotes the volume of the unit round $S^{D-2}$. Since $Q_e/q$ is some numerical factor, we shall not always distinguish $Q_e$ and $q$ as the electric charge.  The neutral solutions were obtained in
\cite{Boulware:1985wk,Cai:2001dz}.

For sufficiently large mass, the solution describes a black hole with both inner and outer horizons, $0\le r_-\le r_+$.  We use the notation $r_0$ to denote a generic horizon, and the corresponding temperature and entropy are
\bea
T&=& \fft{(D-3) r_0^2 \left(r_0^{2 D}-q^2 r_0^6\right)+\alpha  (D-5) r_0^{2 D}}{4 \pi  \left(2 \alpha +r_0^2\right)}\,,\nn\\
S&=&\ft14 \Omega_{D-2} r_0^{D-2} \Big(1+\frac{2 \alpha  (D-2)}{(D-4) r_0^2}\Big)\,.
\eea
It is easy to verify that the first law of black hole thermodynamics $dM=TdS + \Phi(r_0) dQ_e$ is satisfied for both inner and outer horizons.  For given charge $q$, there is a smallest horizon radius $r_{\rm ex}>0$, corresponding to the extremal black hole; it is determined by
\be
\mu_{\rm ex}= r_{\rm ex}^{D-3}+ \frac{\alpha  (D-4) r_{\rm ex}^{D-5}}{D-3}\,,\qquad q^2 =\left(r_{\rm ex}^2 + \fft{D-5}{D-3} \alpha\right)r_{\rm ex}^{2(D-4)}\,.\label{extremal}
\ee
For $\mu> \mu_{\rm ex}$, the black holes have two horizons. If we regard the Euler integrand $E^{(4)}$ as matter, then from the Einstein gravity point of view, the charged black holes satisfy the weak energy condition. It turns out that $\rho - p_{\rm sphere}$ can be negative, which stops the solution from satisfying the dominant energy condition. Furthermore, the trace of the energy-momentum tensor can be positive.

\subsection{Photon spheres and shadows}

Owing to the spherical symmetry, the null geodesic motions can be easily analysed.  For the metric given in (\ref{sol}), the radius of the photon sphere is determined by
\be
\fft{d}{dr}\left(\fft{f}{r^2}\right)\Big|_{\rm r_{\rm ph}}=0\,.\label{pheq}
\ee
The impact parameter, also called the optical radius or the shadow radius is given by
\be
R_{\rm sh} = \fft{r_{\rm ph}}{\sqrt{f(r_{\rm ph})}}\,.
\ee
Note that the form of both radii are independent of the spacetime dimensions. In order to establish (\ref{dconjecture}), it is convenient to define
\bea
{\cal X} &=& \sqrt{\fft{D-1}{D-3}} \fft{R_M}{R_{\rm sh}}\,,\qquad
R_M=\left(\fft{8\pi M (D-1)}{(D-2)\Omega_{D-2}}\right)^{\fft{1}{D-3}}\,,\nn\\
{\cal Y}&=& \sqrt{\fft{D-3}{D-1}} \fft{R_{\rm sh}}{r_{\rm ph}}\,,\qquad {\cal Z} = \left(\fft{2}{D-1}\right)^{\fft{1}{D-3}}\,\fft{r_{\rm ph}}{r_+}.\label{XYZdef}
\eea
Our goal in this paper is to prove
\be
{\cal X}\ge 1\,,\qquad {\cal Y}\ge 1\,,\qquad {\cal Z}\ge 1\,,\label{XYZineq}
\ee
for charged black holes in EGBM gravities in general dimensions.

\subsection{RN black holes}

We begin with analysing the RN black hole in general dimensions by setting $\alpha=0$.  It was stated in \cite{Lu:2019zxb} that (\ref{dconjecture}) was satisfied by these black holes, but no detail was given.  We thus present the proof of this simpler example before we progress to the general solutions. The metric function
is
\be
f=1 - \fft{2\mu}{r^{D-3}} + \fft{q^2}{r^{2(D-3)}}\,.
\ee
The solution describes a black hole when $q\le \mu$, with the outer horizon radius
\be
R_+=\left( \mu + \sqrt{\mu^2 -q^2}\right)^{\fft{1}{D-3}}\,.\nn
\ee
The photon sphere and black hole shadow radii are
\bea
R_{\rm ph} &=& \left(\ft12 (D-1)\mu + \ft12\sqrt{(D-1)^2\mu^2 - 4(D-2) q^2}\right)^{\fft1{D-3}}\,,\nn\\
R_{\rm sh} &=& \frac{2^{-\frac{D-1}{2 (D-3)}} \left((D-1) \mu +\sqrt{(D-1)^2 \mu ^2-4 (D-2) q^2}\right)^{\frac{D-2}{D-3}}}{\sqrt{D-3} \sqrt{(D-1) \mu ^2-2 q^2+\mu  \sqrt{(D-1)^2 \mu ^2-4 (D-2) q^2}}}\,.
\eea
To verify the inequalities (\ref{XYZineq}), it is instructive to introduce a dimensionless parameter $\lambda$ to replace the charge parameter $q$:
\be
q=\fft{\mu \sqrt{\lambda(\lambda + D-1)}}{\lambda + D-2}\,.
\ee
The range $0\le q \le \mu $ is now mapped to $\lambda \in [0,+\infty]$, with $\lambda=0$ giving the Schwarzschild
black hole, and $\lambda=1$ yielding the extremal RN black hole.  We find that
\bea
{\cal X}&=& \fft{(D-1)^{\frac{D-1}{2 (D-3)}} (\lambda+D -2)^{\frac{1}{D-3}} \sqrt{(D-3) \lambda +(D-2)^2}}{(D-2)^{\frac{D-2}{D-3}} (\lambda+D -1)^{\frac{D-1}{2 (D-3)}}}\,,\nn\\
{\cal Y}&=& \Big(1+\frac{\lambda }{(D-1) \left((D-3) \lambda +(D-2)^2\right)}\Big)^{\fft12}\,,\nn\\
{\cal Z}&=& \Big(1 + \frac{(D-3)^2 \lambda/(D-1)}{2 (D-2) \sqrt{(D-3) \lambda +(D-2)^2}+(D-3) \lambda +2 (D-2)^2}\Big)^{\fft{1}{D-3}}.
\eea
The inequalities ${\cal Y}\ge 1$ and ${\cal Z}\ge 1$ are manifest.  The inequality ${\cal X}\ge 1$ can be established by a numerical plot for given $D$. In general $\cal (X,Y,Z)$ are all monotonically increasing functions of $\lambda$.  Near the Schwarzschild limit $\lambda\rightarrow 0$, we have
\bea
\{\cal X,Y,Z\} &=& 1 + \fft{\lambda}{2(D-2)^2}\left\{\frac{1}{D-3},\frac{1}{D-1},\frac{D-3 }{2(D-1)}\right\}
+ {\cal O}(\lambda^2)\,.
\eea
Near the extremal limit $\lambda\rightarrow \infty$, we have
\bea
{\cal X} &=& \frac{(D-2)^{-\frac{D-2}{D-3}} (D-1)^{\frac{D-1}{2 (D-3)}}}{2 \sqrt{D-3}}\Big(2 (D-3)-\frac{1}{\lambda } + {\cal O}(\lambda^{-2})\Big)\,,\nn\\
{\cal Y}&=& \frac{D-2}{2 (D-3)^{3/2} \sqrt{D-1}} \Big(2 (D-3)-\frac{1}{\lambda }+{\cal O}(\lambda^{-2})\Big).\nn\\
{\cal Z}&=& \left(\frac{2(D-2)}{D-1}\right)^{\frac{1}{D-3}}\Big(1 - \fft{1}{\sqrt{(D-3)\lambda}}+
{\cal O}(\lambda^{-1})\Big)\,.
\eea

\subsection{$D=5$ neutral black hole}

We now consider the effect on the inequalities by the Gauss-Bonnet term. It is instructive first to examine a simpler example, namely the neutral black hole in five dimensions:
\be
f=1 + \fft{r^2}{2\alpha}\left( 1 - \sqrt{1 + \fft{8\alpha\mu}{r^{4}}}\right).
\ee
In this case, there is only one horizon $r_+$, determined by
\be
\mu = \ft12 (r_+^2 + \alpha)\,.
\ee
This implies that we must have $\mu> \ft12\alpha$ for the solution to describe a black hole.
The radii of photon sphere and shadow are
\be
r_{\rm ph}=\Big(8\mu(2\mu-\alpha)\Big)^{\fft14}\,,\qquad
R_{\rm sh}=\frac{\alpha \sqrt[4]{8\mu(2 \mu -\alpha) }}{\sqrt{2\mu(2 \mu -\alpha) }-2 \mu +\alpha }\,.
\ee
It is instructive to define a dimensionless parameter $\beta = \alpha/r_+^2$, we then find
\be
{\cal X}= \sqrt{1 + \fft{\sqrt{\beta+1}-1}{\sqrt{\beta+1}+1}}\,,\qquad {\cal Y}= \sqrt{\ft12 (1+\sqrt{\beta+1})}\,,\qquad
{\cal Z}= \sqrt[4]{\beta +1}\,.
\ee
Since the parameter $\beta$ runs from 0 to $+\infty$, it is straightforward to see that the inequalities (\ref{XYZineq}) are all satisfied, with the saturation occurs at $\beta=0$, corresponding to the Schwarzschild black hole.

Since the black hole entropies in higher-order gravities are no longer simply a quarter of the area of the horizon.  The $R_+$ and $M$ relation in (\ref{dconjecture}) is no longer related to the Penrose entropy conjecture.  The black hole entropy can be obtained from the Wald entropy formula, yielding
\be
S=\ft{1}{4}\Omega_3  \left(r_+^3 + 6 \alpha r_+\right)\,.
\ee
The mass and entropy relation now becomes
\be
\fft{256 \pi^3 M^3}{27\Omega_3\, S^2} = \fft{(1 + \beta)^3}{(1 + 6\beta)^2}\,.
\ee
Thus for small but non-vanishing $\beta$, the Penrose conjecture is violated, but it is restored for sufficiently large $\beta$. We may define the effective radius associated with the entropy by $S=\fft14 \Omega_3(\bar R_+^S)^3$, and we have
\be
\bar R^S_+ = \sqrt{r_+^3 + 6\alpha r_+}\,.
\ee
We then have
\be
\fft{R_{\rm ph}}{\sqrt2\, \bar R_{+}^S} = \frac{\sqrt[4]{\beta +1}}{\sqrt[3]{6 \beta +1}}\,.
\ee
Intriguingly, this ratio is a monotonically decreasing function of $\beta$.  In other words, $R_{\rm ph}$ can be smaller than $\bar R_{+}^S$, a clear indication that $R_+$ is a better size parameter than $\bar R_+^S$.

\subsection{General black holes}

The reason we can easily prove the identities in the previous subsections was that we could analytically solve the photon sphere equation (\ref{pheq}) for the photon sphere radius in the RN black holes or the $D=5$ neutral black hole.  This turns out not be possible for the general black holes.  We shall adopt the technique developed in \cite{Lu:2019zxb} to prove the inequalities.  We first prove $Z\ge 1$ and then use this inequality to prove ${\cal X}\ge 1$ and ${\cal Y}\ge 1$. We define a function $W(r)$ as
\be
W(r)=\sqrt{1 + \fft{8\alpha\mu}{r^{D-1}} - \fft{4\alpha q^2}{r^{2(D-2)}}}\, \Big(\fft{f}{r^2}\Big)'\,.
\ee
The photon sphere is located at $r_{\rm ph}$, which is the largest root of $W$.  It can be easily seen that as $r\rightarrow \infty$, $W$ is negative with
\be
W= - \fft{2}{r^3} + \fft{2(D-1)\mu}{r^D} + \cdots
\ee
Since $r_{\rm ph}$ is the largest root, it follows that for any $r$ with $W(r)>0$, then we must have $r<r_{\rm ph}$.  We thus define
\be
\rho=(\ft12 (D-1))^{\fft{1}{D-3}}\, r_+\,,
\ee
and we find
\bea
W(\rho) &=& U - \sqrt{V}\,,\nn\\
U&=&\frac{2}{\rho ^3} + \frac{2^{\frac{D-5}{D-3}} (D-1)^{\frac{2}{D-3}}\alpha}{\rho^5} +\frac{(D-3)^2 q^2}{2\rho^{2D-3}}\,,\nn\\
V&=&\frac{4}{\rho ^6}+\frac{32 \alpha }{(D-1) \rho ^8}+ \frac{ 2^{\frac{5 D-17}{D-3}}\alpha ^2}{(D-1)^{\frac{D-5}{D-3}}\rho ^{10}} + \fft{8(D-3) \alpha q^2}{ \rho ^{2 (D+1)}}\,.
\eea
Note that in the above, we have expressed $\mu$ in terms $r_+$ and hence $\rho$. It is clear that both $(U,V)$ are positive and further more, it is quite straightforward to prove that $U^2 -V\ge 0$ for $\rho>0$.  We therefore demonstrate that $W(\rho)>0$.  It follows that $r_{\rm ph} \ge \rho$, proving that ${\cal Z}\ge 1$.

In order to demonstrate that ${\cal X}\ge1$ and ${\cal Y}\ge 1$, we find it is useful to express $\mu$ in terms of the photon sphere radius $r_{\rm ph}$ by solving (\ref{pheq}).  We have
\bea
\mu &=& \fft{4 \alpha  r_{\rm ph}^{D-5}}{(D-1)^2}+ \fft{(D-2) q^2}{(D-1) r_{\rm ph}^{D-3}}\nn\\
&& + \sqrt{\fft{r_{\rm ph}^{2(D-3)}}{(D-1)^2} + \fft{16 \alpha^2 r_{\rm ph}^{2(D-5)}}{(D-1)^4} + \fft{4(D-3)\alpha q^2}{(D-1)^3 r_{\rm ph}^2} }\,.
\eea
The shadow radius is now given by
\be
R_{\rm sh}=\sqrt{\fft{2\alpha r_{\rm ph}^{2(D+1)}}{(D-2) q^2 r_{\rm ph}^8-(D-1) \mu  r_{\rm ph}^{D+5}+r_{\rm ph}^{2 D} \left(2 \alpha +r_{\rm ph}^2\right)}}\,,
\ee
implying that
\bea
{\cal X}^2 &=& \frac{(D-1)^{\frac{D-1}{D-3}} \mu ^{\frac{2}{D-3}} \left((D-2) q^2 r_{\rm ph}^8-(D-1) \mu  r_{\rm ph}^{D+5}+r_{\rm ph}^{2 D} \left(2 \alpha +r_{\rm ph}^2\right)\right)}{2 \alpha  (D-3) r_{\rm ph}^{2 (D+1)}}\,,\nn\\
{\cal Y}^2 &=& \frac{2 \alpha  (D-3) r_{\rm ph}^{2 D}}{(D-1) \left((D-2) q^2 r_{\rm ph}^8-(D-1) \mu  r_{\rm ph}^{D+5}+r^{2 D} \left(2 \alpha +r_{\rm ph}^2\right)\right)}\,.\label{XYsq}
\eea
The trick now is to make use of ${\cal Z}\ge 1$, which implies
\be
r_{\rm ph} \ge (\ft12 (D-1))^{\fft{1}{D-3}}\, r_+ \ge (\ft12 (D-1))^{\fft{1}{D-3}}\, r_{\rm ex}\,.
\ee
The second inequality holds because $r_{\rm ex}$ is the (smallest) horizon radius of the extremal black hole for given charge $q$, determined by (\ref{extremal}).

We can now define two dimensionless parameters $\beta\ge 0$ and $\gamma> 1$, defined by
\be
\alpha = \beta r_{\rm ex}^2\,,\qquad
r_{\rm ph} = (\ft12 (D-1))^{\fft1{D-3}} \gamma r_{\rm ex}\,.
\ee
Note that the lower bound for $\gamma$ is bigger than 1, but for our purpose, it is sufficient to assume $\gamma>1$.  Substituting $\alpha$ and $r_{\rm ph}$ into (\ref{XYsq}), and we find that both $\cal X$ and $\cal Y$ are functions of the dimensionless quantities $(\beta,\gamma)$ only, with the dimensionful parameter $r_{\rm ex}$ dropped out.  In $D=5$, the expressions are quite simple and they are manifestly no smaller than 1:
\bea
{\cal X}^2 &=& 1 + \frac{\left(4 \gamma ^4-3\right) \left(\sqrt{\beta ^2 \gamma ^2+\beta +4 \gamma ^6}-2 \gamma ^3\right)+\beta  \gamma }{4 \beta  \gamma ^5}\ge 1\,,\nn\\
{\cal Y}^2 &=& 1+ \frac{1+\gamma  \left(\sqrt{\beta ^2 \gamma ^2+\beta +4 \gamma ^6}+\beta  \gamma -2 \gamma ^3\right)}{4 \gamma ^4-1}\ge 1\,.
\eea
For general dimensions, the expressions are much more complicated, we find
\bea
{\cal X} &=& 2^{\frac{2 (D-2)}{(D-3)^2}} (D-1)^{\frac{1-D}{(D-3)^2}} \Big(\fft{\beta}{\gamma^2}\Big) ^{\frac{1}{D-3}}\, \Big(C_1 -\fft{2\sqrt{C_2}}{D-3}\Big)^{\fft12} \Big(C_3 + \sqrt{C_2}\Big)^{\fft1{D-3}}, \nn\\
{\cal Y} &=& \Big(C_1 -\fft{2\sqrt{C_2}}{D-3}\Big)^{-\fft12},
\eea
where
\bea
C_1 &=&1+ \frac{\big(\fft12(D-1)\big)^{\frac{D-1}{D-3}} \gamma ^2 }{(D-3)\beta},\nn\\
C_2&=&1 +\frac{(D-1)^{\frac{5-D}{D-3}} (D-3+(D-5)\beta)}{2^{\frac{2}{D-3}}\beta \gamma ^{2 (D-4)}}
+  \frac{(D-1)^{\frac{2 (D-1)}{D-3}}\gamma ^4}{2^{\frac{4 (D-2)}{D-3}}\beta ^2}\,,\nn\\
C_3 &=& 1+\frac{(D-2) (D-1)^{\frac{5-D}{D-3}} (D-3+(D-5)\beta)}{ (D-3) 2^{\frac{2}{D-3}}\beta \gamma ^{2 (D-4)}}\,.
\eea
A contour plot of variables $(\beta,\gamma)$ can establish the inequalities (\ref{XYZineq}).
In particular, when $\beta =0$, corresponding to the RN black hole in general dimensions, we have
\bea
{\cal X} &=& \frac{1}{2} (D-1) \gamma ^{D-3} \left(1+\frac{4 (D-2)}{(D-1)^2 \gamma ^{2(D-3)}}\right)^{\frac{1}{D-3}} \sqrt{\frac{1}{4} (D-1)^2 \gamma ^{2 (D-3)}-1}\,,\nn\\
{\cal Y} &=& \sqrt{1 + \fft{1}{\fft14 (D-1)^2 \gamma^{2(D-3)} -1}}\,.
\eea
In the limit of $\beta\rightarrow \infty$, $\cal X$ is positive and divergent at the order $\beta^{\fft1{D-3}}$ for $D\ge 6$ and
\be
{\cal Y}= \left(1-\frac{2}{D-3} \sqrt{1+4^{\frac{1}{3-D}} (D-5) (D-1)^{-\frac{D-5}{D-3}} \gamma ^{2 (4-D)}}
\right)^{-\fft12}\,.
\ee
The Schwarzschild black hole limit is achieved by taking $\gamma\rightarrow \infty$.  For large $\gamma$, we have
\be
{\cal X}\sim {\cal Y} = 1+ \frac{2^{\frac{D-1}{D-3}} (D-1)^{\frac{1-D}{D-3}}\beta}{(D-3)\gamma ^2}+\cdots\,.
\ee

\section{Conclusions}

In this paper, we considered charged static black holes in EGBM gravities in general dimensions.  These black holes are spherically symmetric and asymptotic to Minkowski spacetimes. From the view of Einstein gravity, these black holes satisfy the weak energy condition, provided that the Gauss-Bonnet coupling is nonnegative, which also ensures that the perturbation is free of ghost excitations.  There exists an unstable photon sphere outside the horizon, giving rise to the edge of a shadow disk for an observer at infinity. We found the the radii of the horizon, photon sphere and shadow disk satisfy the sequence of inequalities (\ref{dconjecture}), conjectured for the black holes in Einstein gravity.  The robustness of this sequence calls for a better understanding of the underlying condition.

\section*{Acknowledgement}

This work is supported in part by NSFC (National Natural Science Foundation of China) Grant No.~11875200 and No.~11935009.


\begin{thebibliography}{99}

\bibitem{Synge:1966okc}
  J.L.~Synge,
  ``The escape of photons from gravitationally intense stars,''
  Mon.\ Not.\ Roy.\ Astron.\ Soc.\  {\bf 131}, no. 3, 463 (1966).
  doi:10.1093/mnras/131.3.463

\bibitem{Luminet:1979nyg}
  J.P.~Luminet,
  ``Image of a spherical black hole with thin accretion disk,''
  Astron.\ Astrophys.\  {\bf 75}, 228 (1979).

\bibitem{Liu:2019rib}
  H.S.~Liu, Z.F.~Mai, Y.Z.~Li and H.~L\"u,
``Quasi-topological electromagnetism: dark energy, dyonic black holes, stable photon spheres and hidden electromagnetic duality,''
  arXiv:1907.10876 [hep-th].

\bibitem{Lu:2019zxb}
  H.~L\"u and H.D.~Lyu,
  ``On the size of a black hole: the Schwarzschild is the biggest,''
  arXiv:1911.02019 [gr-qc].

\bibitem{Penrose:1973um}
  R.~Penrose,
  ``Naked singularities,''
  Annals N.\ Y.\ Acad.\ Sci.\  {\bf 224}, 125 (1973).
  doi:10. 1111/j.1749-6632.1973.tb41447.x


\bibitem{Hod:2017xkz}
  S.~Hod,
  ``Upper bound on the radii of black-hole photonspheres,''
  Phys.\ Lett.\ B {\bf 727}, 345 (2013)
  doi:10.1016/j.physletb.2013.10.047
  [arXiv:1701.06587 [gr-qc]].

\bibitem{Cvetic:2016bxi}
  M.~Cveti\v c, G.W.~Gibbons and C.N.~Pope,
  ``Photon spheres and sonic horizons in black holes from supergravity and other theories,''
  Phys.\ Rev.\ D {\bf 94}, no. 10, 106005 (2016)
  doi:10.1103/PhysRevD.94.106005
  [arXiv:1608.02202 [gr-qc]].

\bibitem{Mars:2009cj}
  M.~Mars,
  ``Present status of the Penrose inequality,''
  Class.\ Quant.\ Grav.\  {\bf 26}, 193001 (2009)
  doi:10.1088/0264-9381/26/19/193001
  [arXiv:0906.5566 [gr-qc]].

\bibitem{Hod:2012nk}
  S.~Hod,
  ``The fastest way to circle a black hole,''
  Phys.\ Rev.\ D {\bf 84}, 104024 (2011)
  doi:10.1103/PhysRevD.84.104024
  [arXiv:1201.0068 [gr-qc]].

\bibitem{Rasheed:1995zv}
  D.~Rasheed,
``The rotating dyonic black holes of Kaluza-Klein theory,''
  Nucl.\ Phys.\ B {\bf 454}, 379 (1995)
  doi:10.1016/0550-3213(95)00396-A
  [hep-th/9505038].

\bibitem{Lu:2013ura}
  H.~L\"u, Y.~Pang and C.N.~Pope,
  ``AdS dyonic black hole and its thermodynamics,''
  JHEP {\bf 1311}, 033 (2013)
  doi:10.1007/JHEP11(2013)033
  [arXiv:1307.6243 [hep-th]].

\bibitem{Gallo:2015bda}
  E.~Gallo and J.R.~Villanueva,
  ``Photon spheres in Einstein and Einstein-Gauss-Bonnet theories and circular null geodesics in axially-symmetric spacetimes,''
  Phys.\ Rev.\ D {\bf 92}, no. 6, 064048 (2015)
  doi:10.1103/PhysRevD.92.064048
  [arXiv:1509.07379 [gr-qc]].

\bibitem{Bergshoeff:1989de}
  E.A.~Bergshoeff and M.~de Roo,
  ``The quartic effective action of the heterotic string and supersymmetry,''
  Nucl.\ Phys.\ B {\bf 328}, 439 (1989).
  doi:10.1016/0550-3213(89)90336-2

\bibitem{Boulware:1985wk}
  D.G.~Boulware and S.~Deser,
{\it String generated gravity models,}
  Phys.\ Rev.\ Lett.\  {\bf 55}, 2656 (1985).
  doi:10.1103/PhysRevLett.55.2656

\bibitem{Wiltshire:1985us}
  D.L.~Wiltshire,
 ``Spherically symmetric solutions of Einstein-maxwell theory with a {Gauss-Bonnet} term,''
  Phys.\ Lett.\  {\bf 169B}, 36 (1986).
  doi:10.1016/0370-2693(86)90681-7

\bibitem{Cvetic:2001bk}
  M.~Cveti\v c, S.~Nojiri and S.D.~Odintsov,
``Black hole thermodynamics and negative entropy in de Sitter and anti-de Sitter Einstein-Gauss-Bonnet gravity,'' Nucl.\ Phys.\ B {\bf 628}, 295 (2002)
  doi:10.1016/S0550-3213(02)00075-5
  [hep-th/0112045].

\bibitem{Cai:2001dz}
  R.G.~Cai,
{\it Gauss-Bonnet black holes in AdS spaces,}
  Phys.\ Rev.\ D {\bf 65}, 084014 (2002)
  doi:10.1103/PhysRevD.65.084014
  [hep-th/0109133].

\end{thebibliography}
\end{document}